# Degenerate doping in β-Ga$_2$O$_3$ Single Crystals through Hf-doping

Muad Saleh[1,2], Joel B. Varley[3], Jani Jesenovec[1,2], Arkka Bhattacharyya[4], Sriram Krishnamoorthy[4], Santosh Swain[2,5], Kelvin Lynn[1,2,5,6]

[1]Materials Science & Engineering Program, Washington State University, Pullman, WA, 99164, United States of America

[2]Institute for Materials Research, Washington State University, Pullman, WA, 99164, United States of America

[3]Lawrence Livermore National Laboratory, Livermore, CA, 94550, United States of America

[4]Electrical and Computer Engineering, The University of Utah, Salt Lake City, UT, 84112, United States of America

[5]School of Mechanical and Materials Engineering, Washington State University, Pullman, WA, 99164, United States of America

[6]Department of Physics, Washington State University, Pullman, WA, 99164, United States of America

E-mail: kgl@wsu.edu; muad.saleh@wsu.edu

## Abstract

n-type conductivity of β-Ga$_2$O$_3$ grown from the melt is typically achieved using Sn and Si. In this paper, we experimentally and computationally investigate Hf doping of β-Ga$_2$O$_3$ single crystals using UV-Vis-NIR absorption and Hall Effect measurements and hybrid functional calculations. Unintentionally-doped and Hf-doped samples with a nominal concentration of 0.5at% were grown from the melt using vertical gradient freeze (VGF) and Czochralski method in mixed Ar+O$_2$ atmosphere. We demonstrate Hf dopants, predicted to incorporate on the octahedral Ga$_{II}$ site as a shallow donor, achieve degenerate doping in β-Ga$_2$O$_3$ with a measured electron concentration ~2x10$^{19}$cm$^{-3}$, mobility 80-65cm$^2$/Vs, and resistivity down to 5mΩ-cm in our samples. The concentration of Hf was measured to be 1.3x10$^{19}$ atoms/cm$^3$ using glow discharge mass spectroscopy (GDMS) on doped samples, confirming Hf to be the cause of n-type conductivity (electron concentration ~2x10$^{19}$cm$^{-3}$).

## 1. Introduction

β-Ga$_2$O$_3$ is the most thermodynamically stable polymorph of Ga$_2$O$_3$ with monoclinic crystal structure and has been the focus of increasing attention and development in recent years for Schottky diodes, high breakdown voltage switching devices, solar- and visible-blind photodetectors, and gas sensing devices [1-5]. The rapid development has been aided by the availability of β-Ga$_2$O$_3$ single crystals by Czochralski (CZ), Edge-defined fed-film (EFG), vertical gradient freeze (VGF), float zone (FZ), vertical Bridgman (VB), and Verneuil methods [1-10]. β-Ga$_2$O$_3$ is of interest due to being ultrawide-bandgap semiconducting oxide with bandgap of 4.5-4.9eV depending on the polarization and crystal orientation, and tunable electrical conductivity [1-7, 11].

Insulating β-Ga$_2$O$_3$ has been achieved by oxygen annealing or by doping with Fe or Mg which act as deep acceptors [12, 13]. Unintentionally doped (UID) β-Ga$_2$O$_3$ is typically n-type conducting with free electron density of over 1x10$^{17}$cm$^{-3}$, which is attributed to Si impurities [14, 15]. In addition to Si, tunable n-type conductivity in β-Ga$_2$O$_3$ has been typically achieved with Sn and Ge, and also, but less studied, with Ta and Nb; free electron density ($n$) ranging between 1x10$^{17}$cm$^{-3}$ to 2x10$^{19}$cm$^{-3}$, mobility ($\mu$) decreasing with increasing $n$ and ranging between 130cm$^2$/Vs to 25cm$^2$/Vs at room temperature, resistivity ($\rho$) down to 5mΩ-cm, and activation energy (E$_d$) decreasing with increasing $n$ between 7meV to 30meV has been realized with these dopants [3, 6, 14, 16-23].

Si and Sn doping in bulk crystals have been limited to $2 \times 10^{19} cm^{-3}$ due to the second phase formation and crystal quality degradation with Si doping [16], and high evaporation rate of Sn [13, 14]. Ge-doped β-$Ga_2O_3$, to our knowledge, has never been grown as bulk single crystals due to the high vapor pressure of Ge.

Recently, we have achieved tunable n-type conductivity of bulk β-$Ga_2O_3$ with Zr doping and obtained favorable properties for $n$ ($6.5 \times 10^{17} cm^{-3}$ to $5 \times 10^{18} cm^{-3}$), room temperature μ ($112 cm^2/Vs$ to $75 cm^2/Vs$), ρ (down to $17 m\Omega cm$), and $E_d$ (~13meV to below 4meV) [24]. Our analysis revealed that Zr showed a strong preference to substitute for Ga on the octahedral site ($Ga_{II}$), and our initial results identified that Zr doping shows promise for obtaining improved mobilities compared to other dopants in the regime of degenerate doping. Here we explore analogous dopants from the group $d^2$ block of the periodic table, and identify Hf as an additional shallow donor dopant candidate in β-$Ga_2O_3$. We demonstrate both experimentally and theoretically that Hf dopants act as a soluble shallow donor that can be used to achieve degenerate doping in bulk crystals with mobility comparable to or better than those achieved with Zr doping.

## 2. Experimental

UID and 0.5at% Hf-doped β-$Ga_2O_3$ (Hf:β-$Ga_2O_3$) were grown from the melt similar to what has been previously published [24]. 5N purity β-$Ga_2O_3$ source powder was used for all samples. For the Hf-doped samples, the same source powder was mixed for 18 hours in a ball mill at 50rpm with 0.5at% 3N pure $HfO_2$ powder; the 0.5at% is with respect to $Ga_2O_3$ (*i.e.* 0.25at% with respect to Ga). For both growths, two charges were prepared 400-450g each, pressed at 20kpsi, then calcined in an alumina crucible lined with Pt foil at 1500°C for 15 hours. The first charge was first melted, then the second charge was added to increase the volume of the molten material. The charges were melted, and the crystals were then grown in 70mm height by 70mm diameter Iridium crucible by a 25KHz radio frequency inductive heating coil. Ar+$O_2$ gas mixture was used during the melting and the growth with $O_2$<0.2% below 1200°C to limit solid $IrO_2$ formation, $O_2$ ~3%-5% between 1200°C and initial melting of $Ga_2O_3$ to reduce the decomposition and evaporation of $Ga_2O_3$, and 10%-12% during the melting and growth. The $O_2$ content was dropped again to 3% during cooling, and to <0.1% below 1200°C. The Hf:β-$Ga_2O_3$ was grown by unseeded VGF in the Iridium crucible by cooling over the period of 24-30 hours. UID crystals were grown by both CZ using a pull rate of 2mm/hr and 2rpm rotation and VGF with no rotation; the CZ crystal was spiral. Samples for UID crystal were obtained from both the VGF and CZ pulled crystals.

All the tested crystals were obtained by cleaving along the (100) plane; the samples surface had the (100) orientation as confirmed by electron back scattered diffraction (EBSD) and Laue diffraction measurements, as shown in the Supplementary Material. Optical absorption in the UV-Vis-NIR region between 200-3300nm was measured using Cary5 instrument to evaluate the bandedge and sub-bandgap absorption.

Ohmic contacts of 50wt%Ga-50wt%In were obtained by placing them on a freshly cleaved surface followed by annealing at 950°C for 15 minutes; this procedure has been shown previously to yield ohmic contacts down to ~20K [24]. Temperature dependent Hall Effect measurements using 0.51 T magnetic field were performed between 80K and 320K with 10K temperature step using Ecopia HMS 7000 using the Ga-In contacts in van der Pauw configuration.

## 3. Theoretical Assessment

To theoretically assess the solubility and electrical behavior expected for Hf incorporated as a function of growth and doping conditions, we computed the defect formation energies ($E^f$) within a supercell approach, as previously performed to assess Zr dopants [25]. We adopt the exact same computational procedure as in [24, 26, 27], where all defects were modeled within 160-atom supercell representations of bulk β-$Ga_2O_3$, and total energies were computed using the Heyd-Scuseria-Ernzerhof screened hybrid functional (HSE06) as implemented in the VASP code [28-30]. The atomic potentials were treated within projector-augmented wave (PAW) [31] approach and included the $4s^2\ 3d^{10}\ 4p^1$ (Ga), $3p^6 4s^2 3d^2$ (Ti), $4s^2 4p^6 5s^2 4d^2$ (Zr), and $5p^6 6s^2 5d^2$ (Hf) orbitals as valence electrons. The fraction of Hartree-Fock mixing was set to 32%, which yields band gaps and lattice properties in good agreement with experimental values [26]. The solubility-limiting phases considered for the defect formation energies of all extrinsic dopants were taken from the calculated formation enthalpies of the binary-oxide rutile phases of $TiO_2$ ($\Delta H=-9.68$ eV), $ZrO_2$ ($\Delta H=-10.60$ eV), and $HfO_2$ ($\Delta H=-11.29$ eV) using the same Hartree-Fock mixing.

## 4. Results and Discussion

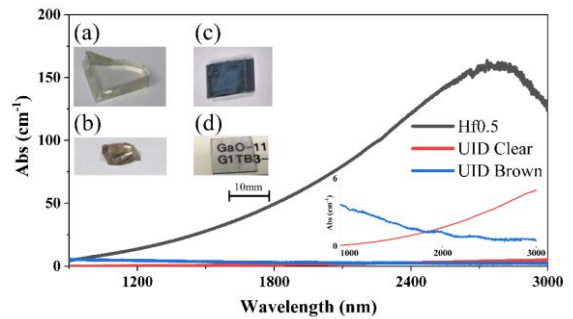

**Figure 1**. NIR absorption of Hf-doped and UID β-$Ga_2O_3$. The left inset show (a) clear yellow UID crystal, (b) clear brown UID crystal (c,d) show clear blue 0.5at% Hf-doped crystal with different thicknesses. The right inset shows a zoomed view of the absorption of the UID samples between 0-6 $cm^{-1}$ with the same x-axis scale.

UID crystals were mostly clear to slightly yellow in color (Figure 1(a)), but UID samples obtained from the bottom of the crucible were slightly brown in color (Figure 1(b)). The Hf samples, similar to Zr doped, were blue in color but transparent and clear (Figure 1(c,d)). NIR absorption spectra is shown in Figure 1. Free carrier absorption can be seen to increase with Hf doping. The UID samples vary slightly in their free carrier absorption in the NIR. The spectrum in the UV-VIS region did not show any additional absorptions and hence is not shown here.

Hall measurements on twelve Hf-doped samples obtained from different locations from the same boule show the electron concentration at room temperature ($n$) to vary between $7.5 \times 10^{18}$ cm$^{-3}$ to $1.9 \times 10^{19}$ cm$^{-3}$ and resistivity between 12.5 mΩ-cm to 5.2 mΩ-cm; eight of the Hf-doped samples are shown as example in Table 1. The variation shows that $n$, and in-turn the Hf concentration, increases vertically toward the bottom of the boule indicating a segregation coefficient of less than 1, and decreases radially toward the edges of the boule indicating a possible convex solid/liquid interface; this interface increases the tendency of spiral growth, and the interface is also typical for highly doped and conductive β-Ga$_2$O$_3$ due to free carrier absorption. The electron concentration achieved with Hf doping is higher than that reached with Zr for the same nominal addition of doping (0.5at%) in the melt. On the basis that Hf incorporates as a shallow donor, this suggests that Hf may be more readily incorporated than Zr, which we find consistent with theoretical predictions as described below. XRD on the lowest part of the grown Hf-doped boule, which is expected to have the highest amount of Hf, did not show any signs of secondary phases; this indicates a possibility that even higher doping of Hf could be achieved.

**Table 1.** Electron concentration, mobility, and resistivity of the 0.5at% Hf-doped and UID β-Ga$_2$O$_3$. Measured at room temperature.

| Sample | Electron Concentration (cm$^{-3}$) | Mobility (cm$^2$/V·s) | Resistivity (mΩ·cm) |
|---|---|---|---|
| Hf-1 | $7.5 \times 10^{18}$ | 78 | 10.7 |
| Hf-2 | $8.2 \times 10^{18}$ | 77 | 10 |
| Hf-3 | $9.2 \times 10^{18}$ | 80 | 8.5 |
| Hf-4 | $9.5 \times 10^{18}$ | 72 | 9.1 |
| Hf-5 | $1.1 \times 10^{19}$ | 70 | 8.1 |
| Hf-6 | $1.1 \times 10^{19}$ | 61 | 9.3 |
| Hf-7 | $1.5 \times 10^{19}$ | 51 | 8.2 |
| Hf-8 | $1.9 \times 10^{19}$ | 64 | 5.2 |
| UID-1 | $4.3 \times 10^{17}$ | 68 | 214 |
| UID-2 | $1.5 \times 10^{17}$ | 97 | 430 |
| UID-3 | Insulating | | |

The Hf-doped samples show higher $n$ than the UID samples which vary between $4.5 \times 10^{17}$ cm$^{-3}$ to semi-insulating. The lower n in the UID samples confirm that Hf is the cause of the conductivity in the Hf-doped samples. The electron concentration in UID samples is possibly due to Si residual impurities and this has been observed by others [14, 15]. The semi-insulating samples were obtained by the VGF boule close to the bottom edge of the crucible and showed a brownish color which could potentially be due to Fe impurities.

Fe impurities, which would tend to be higher at the bottom of the boule, have been measured as an impurity in our raw Ga$_2$O$_3$ powder as shown in glow-discharge mass spectroscopy measurements done by EAG and shown in the Supplementary Material. Fe concentration is $\sim 8.2 \times 10^{16}$ atoms/cm$^3$ in the raw powder used, and Si concentration is $\sim 8.3 \times 10^{17}$ atoms/cm$^3$. Sn and Ge, the other possible suspects for n-type conductivity in β-Ga$_2$O$_3$ were not detected in our raw materials, and are not typically measured impurities in β-Ga$_2$O$_3$ unless they are intentionally added. Further, neither Si, Sn, nor Ge has been shown to show degenerate doping with $2 \times 10^{19}$ cm$^{-3}$ free electron concentration as observed here; which suggests Hf is most likely responsible for the observed carrier concentration. Secondary Ion Mass spectroscopy (SIMS), was not possible to quantify Hf concentration due to the lack of standards, but the Hf existence has been confirmed with energy dispersive x-ray spectroscopy (EDS) in our Hf doped samples as shown in the Supplementary Material. Finally, GDMS on the Hf-doped samples has been conducted and is shown in the Supplementary Materials; the GDMS results show Hf of $1.3 \times 10^{19}$ atoms/cm$^3$ which is close to the measured $n$ and confirm the incorporation of Hf, and show Si, Zr, and Fe to be $2.5 \times 10^{17}$ atoms/cm$^3$, $2.1 \times 10^{17}$ atoms/cm$^3$, and $8.3 \times 10^{16}$ atoms/cm$^3$, respectively. Sn and Ge were below the detection limits.

Temperature dependent Hall effect measurements show degenerate doping in the Hf-doped samples as shown in Figure 2(a); $n$ remains constant with decreasing temperature which has not been shown with bulk single crystalline β-Ga$_2$O$_3$, whereas the UID samples show a decrease in $n$ with decreasing temperature and activation energy between 10.8 meV to 13.8 meV. Mobility μ, as shown in Table 1, in the Hf-doped samples ranges between 80 cm$^2$/Vs and 50 cm$^2$/Vs which is found to generally, but not necessarily, decrease with increasing $n$. The achieved μ is higher than reported for other dopants for the same doping level, and higher than that achieved with our previous Zr doping results [24]. The UID crystals show lower μ for an even lower $n$ than both Zr and the Hf-doped crystals confirming that a different impurity, possibly Si along with compensation from Fe impurities is the cause of n-type doping in our UID crystals. Fig. 2(a,c) shows the carrier concentration, resistivity and mobility as a function of temperature for representative Hf and UID crystals. The samples are expected to have similar structural quality (Hf, UID, and Zr) as they were produced by the same method and procedure, and all the samples were taken from twin free regions (observed by cross-polarized light). For the same reason, it is quite difficult to compare the mobilities presented here to those in the literature, as the structural quality are not always available in the literature, and they were not fully evaluated in the samples presented here.

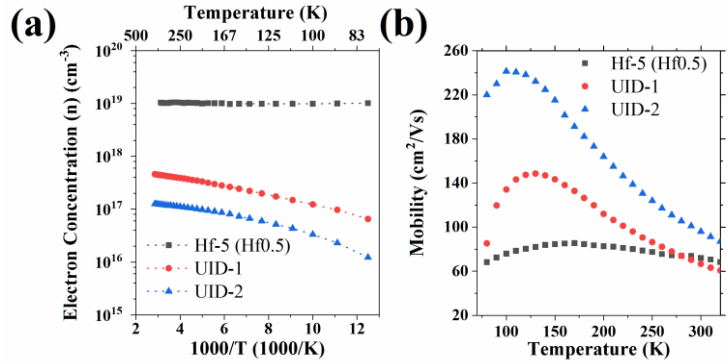

**Figure 2.** Temperature dependent electron concentration and mobility of representative 0.5at% Hf-doped and UID crystals.

These Hall effect measurements were confirmed with another set of independent measurements performed using a Lakeshore 7507 system between 12-320K with Indium contacts; $n$ remained constant to 12K. $n$ were also confirmed with quantum magnetoconductivity measurements (QMC); the low temperature Hall effect and QMC results will be the subject of a different paper.

In Figure 3, we show the calculated defect formation energy diagram for Hf and Zr dopants, for Ga-rich and O-rich growth extremes. Our results identify Hf and Zr behave similarly, most favorably incorporating on the octahedral $Ga_{II}$ site and behaving as shallow donors. Hf is found to exhibit slightly lower formation energies than Zr in both O-rich and Ga-rich extremes, with more Ga-rich conditions leading to higher solubilities owing to less competition with $ZrO_2$ and $HfO_2$ phases. Both dopants also behave as shallow donors on the tetrahedral $Ga_I$ site, but have far higher formation energy and not expected to appreciably incorporate on these sites. Thus Hf is expected to be a more soluble donor dopant than Zr for all conditions, consistent with the realization of higher carrier concentrations in the Hf-doped single crystals as compared to Zr-doped samples for the same nominal doping concentrations (0.5at%) [24].

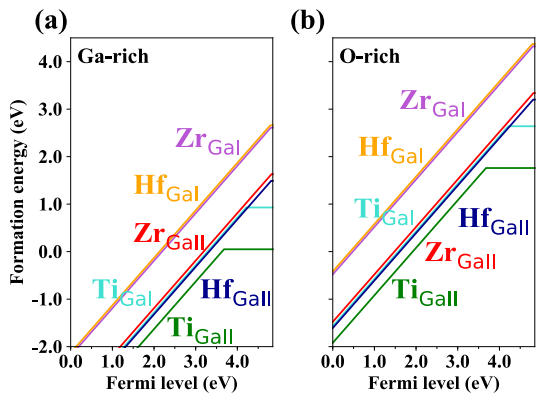

**Figure 3.** Defect formation energies for Hf, Zr, and Ti substitutional dopants in β-Ga2O3, shown for (a) Ga-rich and (b) O-rich growth extremes. $Hf_{Ga}$ shallow donors are found to be more soluble than $Zr_{Ga}$ dopants.

By analogy with Zr and Hf, we also calculated the behavior of $Ti_{Ga}$ as a candidate donor dopant to see if it also exhibited similar electronic behavior. In contrast to Hf and Zr, $Ti_{Ga}$ dopants lead to deep, rather than shallow donor behavior, with a (+/0) transition level 1.18 eV below the conduction band edge (3.37 eV above the VBM) for the more favorable $Ti_{GaII}$. This is due to the higher lying position of the $d$ states associated with the dopants going down the group [32], with the lowest-lying Ti $3d$ states leading to deep states within the band gap. This is supported by analysis of the density of states of the Zr and Hf dopants, whose lowest $d$-derived states for the donor configurations fall 1.65 eV and 2.25 eV above the conduction band minimum, respectively.

## 5. Conclusion

In conclusion, we showed that Hf, similar to Zr, behaves as a soluble shallow donor in β-Ga2O3. We demonstrated degenerate doping with Hf, and electron concentration up to $2 \times 10^{19}$ cm$^{-3}$ mobility of 80-50cm$^2$/Vs, and resistivity down to 5mΩ.cm. Also, we confirmed the Hf incorporation with GDMS measurements, which are strongly correlated with the free carrier concentration determined via Hall measurements.

## Acknowledgements

This material is based upon work supported by the Air Force Office of Scientific Research under award number FA9550-18-1-0507 monitored by Dr. Ali Sayir. Any opinions, finding, and conclusions or recommendations expressed in this material are those of the author and do not necessarily reflect the views of the United States Air Force. This work was partially performed under the auspices of the U.S. DOE by Lawrence Livermore National Laboratory under contract DE-AC52-07NA27344, and supported by the Critical Materials Institute, an Energy Innovation Hub funded by the U.S. DOE, Office of Energy Efficiency and Renewable Energy, Advanced Manufacturing Office. We also thank Jonathan Ogle, Prof. Luisa Whittaker-Brooks at the University of Utah for providing access to the Hall equipment used in this work, and Dr. David Look, T.A. Cooper for low temperature Hall effect measurements, and Adrian Chmielewski and Nasim Alem for useful discussions.

## Supplementary Materials

### *Electron Backscatter Diffraction (EBSD) and Energy-Dispersive X-ray Spectroscopy (EDS)*

EBSD and EDS were performed on a representative Hf:β-$Ga_2O_3$ to confirm the crystal orientation of the sample surface and orientations parallel to (or directions perpendicular to) the sample edges, and the existence of Hf, respectively. The measurements were performed using FEI SIRION scanning electron microscopy (SEM). As can be seen in Figure S1, the sample surface is the (100) plane, and the two other cut/cleaved sample surfaces are the [010] direction and the (001) plane. Although the surface is shown to be ~3.6° from the (100) plane, this is within the deviation range for the measurement. The EDS on the Hf-sample can be seen in Figure S2. Ir is an expected impurity as an Ir crucible is used. As EDS does not have high sensitivity to trace elements (Ir and Hf), the sample measured with EDS were from the bottom of the crucible, which is expected to have higher amounts of Hf, but these samples did not have high optical quality, and were only used with EDS.

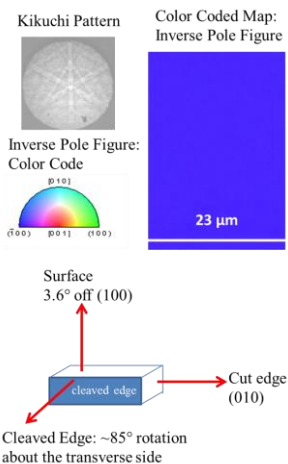

**Figure S1**. EBSD of a representative Hf:β-$Ga_2O_3$ sample showing the Kikuchi pattern, a color coded inverse pole figure map with the color code of 23µm sized area of the sample confirming the sample surface to be (100), and a schematic showing the orientation of the cut [010] side and the cleaved (001) side of the sample.

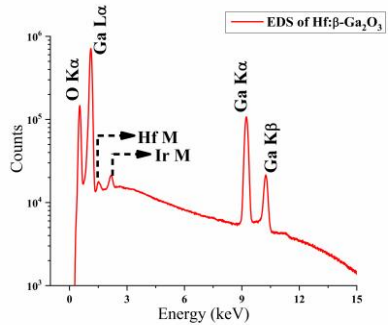

**Figure S2.** EDS of a representative Hf:β-$Ga_2O_3$ sample confirming the incorporation of Hf dopant.

### *Laue Diffraction*

Laue diffraction was conducted at 5N Plus Semiconductors (Utah, USA), to confirm the orientation of the sample surface with a better accuracy; as shown in Fig. S3, the sample surface is within 0.3° from the (100) plane which is within the deviation of the measurement set-up.

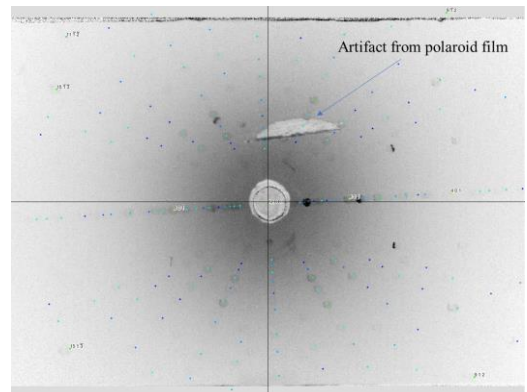

**Figure S3.** Laue diffraction showing the orientation of the surface 0.3° off the (100) plane toward the (010) plane.

### *Glow Discharge Mass Spectroscopy (GDMS)*

GDMS on the raw powder used in the growth and Hf-doped sample were performed at EAG Laboratory (California, USA), and the results are summarized in Table S1. The results show the detected elements, that were present at concentrations above the detection limit.

**Table S1.** The concentration of impurities in the raw $Ga_2O_3$ powder and a Hf doped crystal measured by GDMS. BDL= below detection limit.

| Element | Concentration | | | |
|---|---|---|---|---|
| | Raw Powder | | Hf-doped Crystal | |
| | ppm wt. | atoms/cm$^3$ | ppm wt. | atoms/cm$^3$ |
| B | 0.36 | 1.18x10$^{17}$ | 0.09 | 2.96x10$^{16}$ |
| F | 0.15 | 2.81x10$^{16}$ | BDL | |
| Na | 0.33 | 5.1x10$^{16}$ | 0.33 | 4.01x10$^{16}$ |
| Mg | BDL | | 0.28 | 4.09 x10$^{16}$ |
| Al | 0.45 | 5.93x10$^{16}$ | 21 | 2.76 x10$^{18}$ |
| Si | 6.6 | 8.35x10$^{17}$ | 2 | 2.53 x10$^{17}$ |
| P | BDL | | 0.1 | 1.15x10$^{16}$ |
| S | 1.1 | 1.22x10$^{17}$ | 0.13 | 1.44x10$^{16}$ |
| Cl | 0.41 | 4.11x10$^{16}$ | BDL | |
| Ca | BDL | | 0.4 | 3.55 x10$^{16}$ |
| Ti | 0.05 | 3.70x10$^{15}$ | 0.12 | 8.90x10$^{15}$ |
| Cr | 0.22 | 1.50x10$^{16}$ | 0.47 | 3.21x10$^{16}$ |
| Fe | 1.3 | 8.27x10$^{16}$ | 1.3 | 8.27x10$^{16}$ |
| Ni | BDL | | 0.25 | 1.51x10$^{16}$ |
| Zr | BDL | | 5.4 | 2.1x10$^{17}$ |
| Hf | BDL | | 670 | 1.33x10$^{19}$ |
| W | BDL | | 0.16 | 3.09x10$^{15}$ |
| Ir | BDL | | 3.3 | 6.10x10$^{16}$ |